\documentclass[english,floatfix,twocolumn,showpacs,superscriptaddress,nofootinbib,aps,prd]{revtex4}
\usepackage[T1]{fontenc}
\usepackage{lmodern}
\usepackage{amsmath}
\usepackage{amssymb}
\usepackage{graphicx}
\usepackage{esint}
\usepackage{longtable}
\usepackage{dcolumn}
\usepackage{babel} 
\usepackage{csquotes}

\begin{document}

\title{Constraining the tidal charge of brane black holes using their shadows}

\author{Juliano C. S. Neves}
\email{nevesjcs@if.usp.br}
\affiliation{Centro de Ciências Naturais e Humanas, Universidade Federal do ABC,\\ Avenida dos Estados 5001, Santo André, 09210-580 São Paulo, Brazil}


\begin{abstract}
A constraint on the tidal charge generated within a brane world is shown. Using the shadow of a rotating black hole in 
a brane context in order to describe the M87* parameters recently announced by the Event Horizon 
Telescope Collaboration, the deviation from circularity of the reported shadow produces an upper bound 
on the bulk's nonlocal effect, which is conceived of as a tidal charge in the four-dimensional brane 
induced by the five-dimensional bulk. Therefore, a deviation from circularity $\lesssim 10\%$ leads to an upper bound 
on the tidal charge $\lesssim 0.004M^2$. 
\end{abstract}

\pacs{04.50.-h,11.25.-w,04.80.Cc,11.25.-w}

\maketitle

\section{Introduction}

The recent shadow of the supermassive black hole M87* at the center of the Messier 87 galaxy built by
the Event Horizon Telescope Collaboration (EHT)  \cite{EHT,EHT2} is a new window for studies in strong 
gravitational field. The shadow cast by black holes depends on the metric parameters, and the main candidate 
for the M87* shadow is the Kerr geometry, despite the existence of alternatives to the M87* shadow like 
superspinars \cite{Bambi} and regular black holes \cite{Neves,Gosh}. The shadow reported by the collaboration 
presents a deviation from circularity
less than ten percent, i.e., $\Delta C \lesssim 10\%$.  Using constraints coming from the EHT, researchers have 
obtained upper bounds on the rotation parameter of M87* \cite{Bambi}, extra dimension 
length \cite{Vagnozzi}, cosmological parameters \cite{Yu}, magnetic charge from black holes in
 nonlinear electrodynamics \cite{Khodadi}, and on the parameter of the generalized 
uncertainty principle \cite{Neves}. The importance of the shadow phenomenon is due to use of an appropriate
 geometry for studies on the strong gravitational field,  a geometry that describes a rotating black hole.   

Shadows of objects have been studied since the Synge work \cite{Synge} on the shadow of 
the nonrotating Schwarzschild  black hole. In the same direction, Bardeen \cite{Bardeen2} built
  the first shadow for the Kerr geometry, which is a rotating black hole. Since those pioneer works, shadows of different black holes have been published, like shadows for the Reissner-Nordström
    black hole \cite{Zakharov}, the Kerr-Newman black hole \cite{Vries}, black holes with a cosmological constant
     \cite{Perlick}, Kerr-Newman-NUT black hole \cite{Grenzebach}, Kerr-Newman-Kasuya 
     black hole \cite{Ovgun}, regular black holes \cite{Zi_Bambi,Abdujabbarov,Amir, Neves} and 
     for braneworld black holes \cite{Eiroa1,Eiroa2}. In this work, a vacuum brane will be considered, then
  the role of an accretion disk in the shadow phenomenon will be an issue for a future work. 
  Indeed, models of accretion disks for different geometries are still 
  in development \cite{Perlick2,Cunha,Roman,Herdeiro}.  
     
     Shadows of rotating black holes with a cosmological constant in a brane context will be focused on this work
      in order to describe the M87* data and produce an upper bound on the tidal charge induced by the extra dimension.         
The brane world adopted here is the second Randall-Sundrum model. The Randall-Sundrum
  model one (RS-I)  \cite{Randall-Sundrum} tries to solve the hierarchy problem, namely, the huge difference
   between the electroweak scale
 $(\sim \text{TeV})$ and the four-dimensional Planck scale $(\sim 10^{16} \text{TeV})$. According to the RS-I,
  that problem is solved with introduction of a warp factor in the bulk metric, which is conceived of as a
   five-dimensional asymptotically anti-de Sitter spacetime with two four-dimensional branes as boundaries. 
   One of those branes
    is our four-dimensional universe, and the other one is a hidden universe. In the RS-I, the extra dimension---from
     the bulk---is compact with a finite radius. On the other hand, the second Randall-Sundrum model (RS-II)
      \cite{Randall-Sundrum 2}, focused on this article, assumes only a brane, and the extra dimension radius
       is infinite.\footnote{See reviews on brane worlds in Refs. \cite{Maartens,Clifton}.} 

In order to construct solutions of black holes in a Randall-Sundrum
scenario, it is necessary to consider new four-dimensional gravitational field equations.  Following 
Shiromizu \textit{et al.} \cite{Shiromizu}, induced equations were obtained from the Gauss-Codazzi equations 
in five dimensions. According to the authors, by assuming the RS-II-type model and the $\mathbb{Z}_{2}$ 
symmetry in the bulk, the effective gravitational field equations are obtained on the brane. With those field equations
on the brane, spherically and axially symmetric solutions (black holes with and without rotation, wormholes) were widely
studied \cite{Dadhich_Maartens,Casadio,Bronnikov,Aliev,Molina_Neves,Molina_Neves_2,Neves_Molina,Lemos-2003, Lobo2, Molina3, Prado,Neves2}.

Shadows of rotating black holes in a RS-II-type scenario used here were built by Eiroa and Sendra \cite{Eiroa2} from our
work \cite{Neves_Molina}, where such solutions or geometries were obtained for the first time. In such a 
context, the brane field equations  carry the bulk's influence on the four-dimensional spacetime. 
In regard to rotating black holes in the RS-II-type model, the geometry that
describes those objects presents a tidal charge, commonly conceived of as a nonlocal bulk's influence on the
brane \cite{Maartens,Aliev,Neves_Molina}. Therefore, I will constrain that tidal charge using the shadow 
silhouette and its deformation or deviation from circularity. By adopting the M87* parameters with the
 angle between the black hole rotation axis and the observer position $\theta_{obs}=17^{\circ}$, which is in agreement
  with the observed jets supposedly aligned with the black hole rotation axis \cite{EHT,Walker}, the deviation 
  from circularity reported by the collaboration ($\lesssim 10\%$) provides an upper bound on the tidal charge
   $\lesssim 0.004 M^2$. This result, as I pointed out, is obtained from a rotating black hole. 
   On the other hand, constraints on the tidal charge have been reported using, in general, metrics with 
   spherical geometry, especially a Reissner-Nordström-type geometry (with a tidal charge instead of a 
   Coulomb charge \cite{Bohmer,Kotrlova,Horvath,Zakharov2,Bin-Nun}) that appears in brane contexts as it was 
    reported in Ref. \cite{Dadhich_Maartens}. An exception  is Banerjee \textit{et al.} \cite{Banerjee} in 
    which rotating black holes (without a cosmological constant) are used in the brane context. According to 
    those authors, negative values of the tidal charge are favored with the M87* data. However, positive values
 are not ruled out. In this sense, this work presents an upper bound on the tidal charge (the positive one), and, as we will see, the normalized constraint evaluated here using a rotating black hole with a cosmological constant is the second best found in the literature. 

The structure of this paper is as follows: in Section II, the gravitational field equations and  the spacetime metric that
describes a rotating black hole with a cosmological constant in a brane world are presented. 
The equations to draw the shadow are shown.
 In Section III, it is indicated the procedure in order to evaluate the shadow's deviation from circularity. An 
 upper bound on the tidal charge is obtained from that procedure. The final remarks are made in Sec. IV.\footnote{In this 
 article, one adopts geometrized units in which $G=c=1$, where $G$ is the gravitational constant and $c$ is 
 the speed of light in vacuum. For the evaluation of the tidal charge, the M87* parameters were used, thus the values
 of $G$ and $c$ were restored.}

\section{Shadows of rotating black holes with a cosmological constant in a brane world}
This section deals with a solution of rotating black holes in a brane world with a cosmological constant 
published by us in Ref. \cite{Neves_Molina}, whose shadows were constructed in Ref. \cite{Eiroa2}. 
I intent to point out some features of that shadows to apply them in the next section for the deviation from 
circularity analysis. More details of the geodesics and shadows calculations are found in the two references 
mentioned above.   

\subsection{Brane world context: the gravitational field equations}
In order to build rotating black holes with a cosmological constant in a different realm from general relativity,
 new field equations are necessary. In the brane context,
the Einstein field equations are replaced by field equations projected onto the brane. Then, the brane is conceived of as
a four-dimensional world immersed in a five-dimensional bulk, which is an asymptotically anti-de Sitter
spacetime. The field equations on the brane were elegantly obtained by Shiromizu, Maeda, and Sasaki  \cite{Shiromizu}.
The induced vacuum field equations on the brane are written as
\begin{equation}
G_{\mu\nu} = -\Lambda_{4D}g_{\mu\nu} - E_{\mu\nu},
\label{field_equations}
\end{equation}
in which $G_{\mu\nu}$ is the four-dimensional Einstein tensor related to the brane metric $g_{\mu\nu}$, and
 $\Lambda_{4D}$ is the four-dimensional brane cosmological constant. The tensor $E_{\mu\nu}$ is proportional to
the (traceless) projection of the five-dimensional Weyl tensor onto the brane. It is worth noting that the vacuum
 field equations given by Eq. (\ref{field_equations}) reduce to the vacuum Einstein equations in the low energy limit. 

By using the effective field equations, written without specifying the traceless tensor $E_{\mu\nu}$, the trace of Eq.~(\ref{field_equations}) delivers a constraint on the Ricci scalar given by
\begin{equation}
R=4\Lambda_{4D}.
\label{Ricci}
\end{equation}
In order to solve Eq.~(\ref{Ricci}) and produce rotating black holes, it is assumed the axial symmetry in the
brane indicated by the Kerr-Schild-(anti)-de Sitter \textit{Ansatz} \cite{Gibbons}. Accordingly,
\begin{equation}
ds^{2} = ds_{\Lambda}^{2} + H \left( l_{\mu}dx^{\mu} \right)^{2},
\label{Kerr-Schild}
\end{equation}
where the symbol $ds_{\Lambda}^{2}$ indicates the AdS (anti-de-Sitter) or dS (de Sitter) metric. It is worth emphasizing that the main interest here,  due to cosmological observations, is within the asymptotically dS rotating black holes in the brane. Lastly, the remaining parameters in Eq. (\ref{Kerr-Schild}) are: $H(r,\theta)$ is a function of  the radial and polar coordinates (when spherical coordinates are adopted), and $l_{\mu}$ stands for a null vector field. Substituting the \textit{Ansatz} (\ref{Kerr-Schild}) into the constraint (\ref{Ricci}), one has $H(r,\theta)$ and the complete metric in the
Kerr-Schild form (\ref{Kerr-Schild}). That is to say,  
\begin{equation}
H(r,\theta) = \frac{2Mr}{\Sigma} - \frac{q}{\Sigma},
\label{H}
\end{equation}
with
\begin{equation}
 \Sigma = r^{2} + a^{2} \cos^{2} \theta,
\end{equation}
give us a spacetime with axial symmetry that will be more evident with the aid of the Boyer-Lindquist coordinates. In 
such coordinates, the parameters $M$, $a$, and $q$ will be properly interpreted.

\subsection{Spacetime metric and geodesic equations}
The spacetime metric (\ref{Kerr-Schild}) in  the  Boyer-Lindquist coordinates ($t,r,\theta,\phi$) is given by (details in
Ref. \cite{Neves_Molina})
\begin{eqnarray}
ds^{2} & = & -\frac{1}{\Sigma}\left(\Delta_{r} - \Delta_{\theta} a^{2} \sin^{2} \theta\right) dt^{2} \nonumber \\
&& - \frac{2a}{\Xi\Sigma} \left[(r^{2} + a^{2})\Delta_{\theta} - \Delta_{r}\right] \sin^{2} \theta dt d\phi \nonumber \\
&& + \frac{\Sigma}{\Delta_{r}}dr^{2} + \frac{\Sigma}{\Delta_{\theta}}d\theta^{2}\nonumber \\
& &  + \frac{\sin^{2}\theta}{\Xi^{2}\Sigma}\left[(r^{2}+a^{2})^{2}\Delta_{\theta}-\Delta_{r}a^{2}\sin^{2}\theta\right] d\phi^{2},
\label{Boyer-Lindquist}
\end{eqnarray}
where
\begin{equation}
\Delta_{\theta} = 1 + \frac{\Lambda_{4D}}{3} a^{2} \cos^{2} \theta, \hspace{0.3cm} \Xi = 1 + \frac{\Lambda_{4D}}{3} a^{2},
\label{definitions}
\end{equation}
and
\begin{equation}
\Delta_{r} = (r^{2} + a^{2})\left( 1 - \frac{\Lambda_{4D}}{3}r^{2}\right) - 2Mr + q.
\label{Delta_r}
\end{equation}
From the Boyer-Lindquist coordinates, the parameters $M$, $a$, and $q$ are interpreted as mass parameter, rotation
parameter, and charge parameter, respectively. Indeed, according to Refs. \cite{Caldarelli,Dehghani}, the parameter $M$
of the Kerr-Newman-(A)-dS is related to the mass in asymptotically dS and AdS solutions, in such spacetimes 
the black hole mass is 
$\pm M/\Xi^2$ (plus for AdS and minus for dS). As I pointed out, the brane is assumed empty, a vacuum brane. 
Then, it is common to conceive of $q$ as a tidal charge  (see \cite{Aliev,Neves_Molina,Maartens}), i.e., 
it is a nonlocal bulk's influence on the brane. The charge feature of $q$ is evident once one has the components of $E_{\mu\nu}$:\footnote{There are typos in our work \cite{Neves_Molina} regarding the components of 
$E_{\mu\nu}$.}
\begin{eqnarray}
E_{t}^{t} & = & -E_{\varphi}^{\varphi} = q \left[ \frac{2(r^{2} + a^{2})}{\Sigma^{3}} - \frac{1}{\Sigma^{2}} \right] , \nonumber \\
E_{r}^{r} & = & -E_{\theta}^{\theta} = \frac{q}{\Sigma^2}, \nonumber \\
E_{\varphi}^{t} & = & -\frac{\left(r^{2} + a^{2}\right) \sin^{2}\theta}{\Xi} E_{t}^{\varphi}= -\frac{2qa}{\Xi\Sigma^{3}}\left(r^{2} + a^{2}\right) \sin^{2}\theta. \nonumber \\
\label{E}
\end{eqnarray}
In the brane world, the traceless tensor $E_{\mu\nu}$ is thought of  as a geometric component
of Eq.~(\ref{field_equations}). As I said, such a tensor is a part of the five-dimensional Weyl
tensor projected onto the brane and gives a geometric influence on the vacuum brane. However, Eq. (\ref{field_equations}) may be interpreted as the Einstein field equations with a traceless energy-momentum
tensor as well ($T^{\mu}_{\mu}=E^{\mu}_{\mu}=0$). In the general relativity realm,
 Eq. (\ref{E}) is the Kerr-Newman-(A)-dS energy-momentum tensor since one replaces $q$ by $Q^2$, i.e.,  
\begin{equation}
q \rightarrow -\frac{Q^2}{8\pi},
\label{q_Q}
\end{equation}
with $Q^2$ playing the role of the Coulomb charge. This is the main argument in favor of a tidal charge 
interpretation of $q$. But, contrary to Kerr-Newman-(A)-dS geometries, negative and positive values of  the tidal 
charge present different spacetime structures (at least their scales). On the other hand, in the 
Kerr-Newman-(A)-dS spacetimes, positive and negative values of $Q$ provide the same localization for 
horizons and the ergosphere. Because of the tidal charge is given by $q$ instead of $q^2$, negative values 
of $q$ increase the ergosphere radius compared to positive ones (see 
details in Refs. \cite{Aliev,Neves_Molina}). In this sense, for $q<0$, brane black holes are more appropriate to the Penrose mechanism of extracting energy from the black hole rotation.  

As I am concerned with the cosmological context, the positive cosmological constant will be
 considered  in this article, in agreement with the observations of an accelerating universe \cite{Planck}.
  In this case, following Ref. \cite{Neves_Molina}, regarding $a^2< M^2$ and $q_{min}<q<q_{max}$, the function $\Delta_r$ provides three positive roots, which are null surfaces: the inner horizon $r_-$, the event horizon $r_+$, and the cosmological horizon $r_{++}$.  Therefore, the spacetime structure
  is indicated by
\begin{equation}
r_-<r_+<r_{++}.
\label{Structure}
\end{equation} 
On the other hand, zeros of the metric component $g_{tt}$ provide us Killing surfaces and the limit of the ergosphere.
With the M87* mass parameter $M=(6.5 \pm 0.7 )\times 10^{9}M_{\odot}$ for $a^2<M^2$, i.e., avoiding an 
overspinning black hole, the metric term $g_{tt}$ has one real root in between the event and the cosmological 
horizon ($r=S$). Thus, with this new element, the spacetime structure reads
\begin{equation}
0<r_-<r_+<S<r_{++}.
\label{Structure2}
\end{equation} 
The region $r_+ <r<S$ indicates the ergosphere. 

It is commonly claimed \cite{Dadhich_Maartens,Aliev} that a negative value of the tidal charge
 is more \enquote{natural} since it
 confines the gravitational field near the brane. As we will see, this statement presents a new effect when
  it is considered the black hole shadow. For negative values of $q$, the brane black hole presents a less deformed
   shadow when it is compared to the shadow generated by a positive value of the tidal charge. A clear deviation from circularity is possible only for positive values of the tidal charge. According to previous works, positive values
   of the tidal charge generate more modifications on the brane like turning a spacelike singularity into a timelike
   singularity \cite{Dadhich_Maartens}. Due to this ability, positive tidal charges modify strongly the shadow's shape. 
   
   It is worth noting that the upper bound on the tidal charge obtained in Sec. III, by using the black hole shadow, will 
   be better than $q_{max}$ evaluated in order to allow the spacetime structure given by (\ref{Structure2}).

\subsection{Shadow and its silhouette}

The black hole shadow is drawn from the geodesic equations. Geodesics for a rotating black hole, namely the Kerr
 metric, were obtained by Carter \cite{Carter}, who indicated the separability of the geodesic equations and argued that a test particle in the Kerr  spacetime  presents four constants of motion along the geodesics. Such constants are the two Killing vector fields $\xi_t$ and $\xi_\phi$ with their respective constants, the mass of the test particle along the 
 geodesics, and the last constant is called the Carter constant (in general, it is indicated by $K$). 
 The Kerr-anti-de Sitter and Kerr-de Sitter spacetimes \cite{Hackmann} and the solution with rotation in a brane world adopted in this article present those same four constants for test particles along geodesics \cite{Neves_Molina}. 

From the two Killing vector fields $\xi_{t}$ and $\xi_{\phi}$ indicated directly by the geometry with axial symmetry given by Eq. (\ref{Boyer-Lindquist}), one has the particle's constants of motion $E$ and $L$ along the geodesic, energy and angular momentum, respectively,
\begin{equation}
p_{t}=-E \ \ \  \mbox{and} \ \ \  p_{\phi} = L.
\end{equation}
All the geodesic equations for the spacetime given by Eq. (\ref{Boyer-Lindquist}) were obtained in Ref. \cite{Neves_Molina} using the Hamilton-Jacobi equation. For our purpose, only the radial and azimuthal components
are useful:
\begin{eqnarray}
\Sigma\dot{r} &= &\sqrt{\mathcal{R}}=\sqrt{ P^{2}-\Delta_{r}\left(\pm\delta^{2}r^{2}+K\right)}, \nonumber \\
\Sigma\dot{\phi} &=&  \frac{aP}{\Delta_{r}}-\frac{1}{\Delta_{\theta}}\left[aE-\left(1+\frac{\Lambda_{4D}}{3}a^{2}\right)  \textrm{cosec}^{2}\theta \ L \right], \nonumber \\
\label{Geodesic}
\end{eqnarray} 
with
\begin{equation}
P  = \left(r^{2}+a^{2}\right)E - \left(1+\frac{\Lambda_{4D}}{3}a^{2}\right)a L,
\end{equation}
and dot indicates derivative with respect to an affine parameter. The parameter $\delta$ is the particle's mass (one of
the four constants along the geodesics), and for $\delta=0$, the case studied here, one has null geodesics.

Following \cite{Grenzebach,Eiroa2,Neves}, two new parameters, which are constants in the shadow's silhouette, are defined as
\begin{equation}
\xi=\frac{L}{E} \ \ \ \ \mbox{and} \ \ \ \ \eta=\frac{K}{E^2}.
\label{Constants}
\end{equation}
The shadow silhouette or its form is provided by unstable photon orbits with $r=r_p>r_+$ constant outside the event horizon. For those orbits, photons may eventually either fall into the black hole or escape to the observer
 position $r_{obs}\gg r_p$. Therefore, according to the geodesic equation for the radial coordinate, $\dot{r}$, one has $\mathcal{R}(r_p)=\mathcal{R}'(r_p)=0$ ( in which the symbol $'$ means derivative with respect to $r$) in order to deliver the unstable orbits that draw the shadow silhouette. In general, for black holes with rotation, $r=r_{p-}$ and
 $r= r_{p+}$, where $r_{p-}\leq r_{p+}$, and the black hole shadow is built from an interval of unstable orbits 
 in which $r_{p-}$ is a minimum and $r_{p+}$ is a maximum value. It is worth noting that for the Schwarzschild black hole $r_{p-}=r_{p+}=3M$ and the shadow is circular, with $M$ playing the role of  the Arnowitt-Deser-Misner mass for such a black hole. But due to rotation, the Kerr black hole and Kerr-like black holes do not
 present circular shadows for any set of parameters. As we will see, those shadows are deformed by large black hole
  rotation and, in the brane context, the tidal charge $q$ contributes substantially to the shadow's shape. 
 
The main equations for the shadow calculation that involve the radial coordinate or $\mathcal{R}(r_p)$ and its derivative lead to
\begin{eqnarray}
\eta(r_p) &=& \frac{16r_p^2 \Delta_r(r_p)}{\Delta'_{r} (r_p)^2}, \\
\xi (r_p) &=& \frac{(r_p^2+a^2)\Delta_r'(r_p)-4r_p\Delta_r(r_p)}{\Xi a \Delta'_r(r_p)}. 
\end{eqnarray}
As we will see, these conserved quantities are part of the shadow's equations, are constant in the shadow silhouette.
The constants $\eta(r_p)$ and $\xi (r_p)$ are terms of the celestial coordinates that are used by the observer in order
to draw the shadow. 

As we saw, the geometry studied here is not asymptotically flat, it is asymptotically dS. Therefore, an observer 
is not at infinity describing the shadow. For $\Lambda_{4D}>0$, there is a cosmological horizon in the brane.
   In this sense, the observer is located at the
  so-called \textit{domain of outer communication}, which is the region defined 
   between the event horizon $r_+$ and the cosmological horizon $r_{++}$ in asymptotically dS spacetimes.
    Then, our observer is located at the point with coordinates ($r_{obs},\theta_o$). Because of the axial symmetry,
     the black hole shadow (its silhouette) depends only on both the radial and
      polar coordinates. The coordinate $\theta_o$,  which is the position in which the shadow
       is observed in relation to the rotation axis, stands for the observer angle, and the parameter $r_{obs}$ is the distance  from the black hole. In this article, one assumes the M87* parameters to obtain a constraint on the tidal
       charge, that is to say,  $\theta_o = 17^{\circ}$ and $r_{obs} = (16.8 \pm 0.8)$Mpc.
       As we are dealing with an asymptotically dS spacetime,  it is also adopted the orthonormal tetrad
        $e_a^\mu=(e_0^\mu,e_1^\mu,e_2^\mu,e_3^\mu)$ in order to describe the shadow silhouette
         (see Refs. \cite{Grenzebach,Eiroa2}) in the domain of outer communication, according to the observer
         position. Then, the null congruences coming from
          $r_{p-}\leq r \leq r_{p+}$ reaches the distant observer and are projected onto the tetrad. The shadow 
          phenomenon is described by using $e_a^\mu$, which is written with the aid of the coordinate basis vectors $(\partial_t,\partial_r, \partial_\theta, \partial_\phi)$, that is to say, 
\begin{eqnarray}
e_0&=&\frac{(r^2+a^2)\partial_t + a\Xi \partial_\phi}{\sqrt{\Delta_r\Sigma}}\bigg\vert_{(r_{obs},\theta_o)}, \nonumber \\
e_1&=&\sqrt{\frac{\Delta_\theta}{\Sigma}}\partial_\theta\bigg\vert_{(r_{obs},\theta_o)}, \nonumber 
\end{eqnarray}
\begin{eqnarray}
e_2&=&-\frac{\Xi\partial_\phi+a\sin^2\theta\partial_t}{\sqrt{\Delta_\theta\Sigma}\sin\theta}\bigg\vert_{(r_{obs},\theta_o)}, \nonumber  \\ 
e_3&=&-\sqrt{\frac{\Delta_r}{\Sigma}}\partial_r\bigg\vert_{(r_{obs},\theta_o)}.
\label{Tetrad}
\end{eqnarray}
According to the authors of Ref. \cite{Grenzebach}, the direction of the vector $e_3$ points toward the black hole, and
the component $e_0$ is the observer's four-velocity. Therefore, in this case, the observer is not necessarily at rest. It
is worth mentioning that the orthonormal tetrad is chosen such that $e_0\pm e_3$ is tangential to the principal null congruence direction for a metric with axial symmetry like the metric studied here.

Following Refs. \cite{Grenzebach,Eiroa2}, the light rays that come from the region defined by $r_p$ 
(the shadow's silhouette) are conceived of as curves $\gamma$ or a null congruence such that their tangent vectors in the
coordinate basis are
$\dot{\gamma} =  (\dot{t}\partial_t,\dot{r}\partial_r,\dot{\theta}\partial_\theta,\dot{\phi}\partial_\phi)$. When
projected onto the orthonormal tetrad, $\dot{\gamma}$ reads
\begin{equation}
\dot{\gamma} =  \zeta \left(-e_0+\sin\alpha\cos\beta e_1+\sin\alpha\sin\beta e_2+\cos \alpha e_3  \right),
\label{Gamma_dot}                       
\end{equation}
and that vector is given at the observer's position by using the chosen tetrad 
(\ref{Tetrad}).  The new angles $\alpha$ and $\beta$ are the so-called celestial coordinates, and the description of the 
black hole shadow is entirely made by using these coordinates (see Fig. 1 from Ref. \cite{Neves} for details). 
Indeed, as we will see, the celestial coordinates will be replaced by Cartesian coordinates in order to draw the shadow. 

The factor $\zeta$  and the celestial coordinates $\alpha$ and $\beta$ are straightforwardly obtained 
from Eq. (\ref{Gamma_dot}), the tangent vector $\dot{\gamma} =  (\dot{t}\partial_t,\dot{r}\partial_r,\dot{\theta}\partial_\theta,\dot{\phi}\partial_\phi)$, and the orthonormal tetrad equations, in such a way that
\begin{equation}
\zeta=-\frac{(r^2+a^2)E-a\Xi L}{\sqrt{\Delta_r\Sigma}} \bigg\vert_{(r_{obs},\theta_o)},
\end{equation}  
and, at the same time, 
\begin{eqnarray}
\sin \alpha&=&\frac{\sqrt{\Delta_r \eta(r_p)}}{(r^2+a^2)-a\Xi \xi(r_p)} \bigg\vert_{(r_{obs},\theta_o)}, \\
\sin \beta &=& \frac{\left( \Xi \xi(r_p)\csc^2 \theta-a \right)\sin \theta}{\sqrt{\Delta_\theta \eta(r_p)}} \bigg\vert_{(r_{obs},\theta_o)},
\end{eqnarray}
where the geodesic equations for the components $\dot{r}$ and $\dot{\phi}$, given by Eq. (\ref{Geodesic}), were
used (see details in Ref. \cite{Eiroa2}). It is worth emphasizing that the equations for the celestial coordinates 
depend on the observer's position $(r_{obs},\theta_o)$ and parameters of orbits $r=r_p$ that characterize the
so-called photon \enquote{spheres}.

It is common in articles \cite{Grenzebach,Eiroa2,Neves} that deal with asymptotically dS and AdS black holes to replace the celestial coordinates by Cartesian coordinates in order to describe the shadow's shape. In the Cartesian coordinates, the shadow is drawn by the parametric equations
\begin{eqnarray}
x(r_p)&=&-2\tan\left(\frac{\alpha(r_p)}{2} \right)\sin\left(\beta(r_p) \right), 
\label{x} \\
y(r_p)&=&-2\tan\left(\frac{\alpha(r_p)}{2} \right)\cos\left(\beta(r_p) \right).
\label{y}
\end{eqnarray}
Before constructing the deviation from circularity, it is appropriate to adopt an approximation for the Cartesian coordinates $x$ and $y$, for the value of $\alpha$ is a small quantity due to the large distance of the observer.
 Then, Eqs. (\ref{x}) and (\ref{y}) are rewritten as     
\begin{eqnarray}
x(r_p)&\simeq&-\sin \left(\alpha(r_p)\right)\sin\left(\beta(r_p) \right),
\label{x_approx}  \\
y(r_p)&\simeq&-\sin \left(\alpha(r_p)\right)\cos\left(\beta(r_p) \right).
\label{y_approx}
\end{eqnarray}

\begin{figure}
\begin{centering}
\includegraphics[scale=0.45]{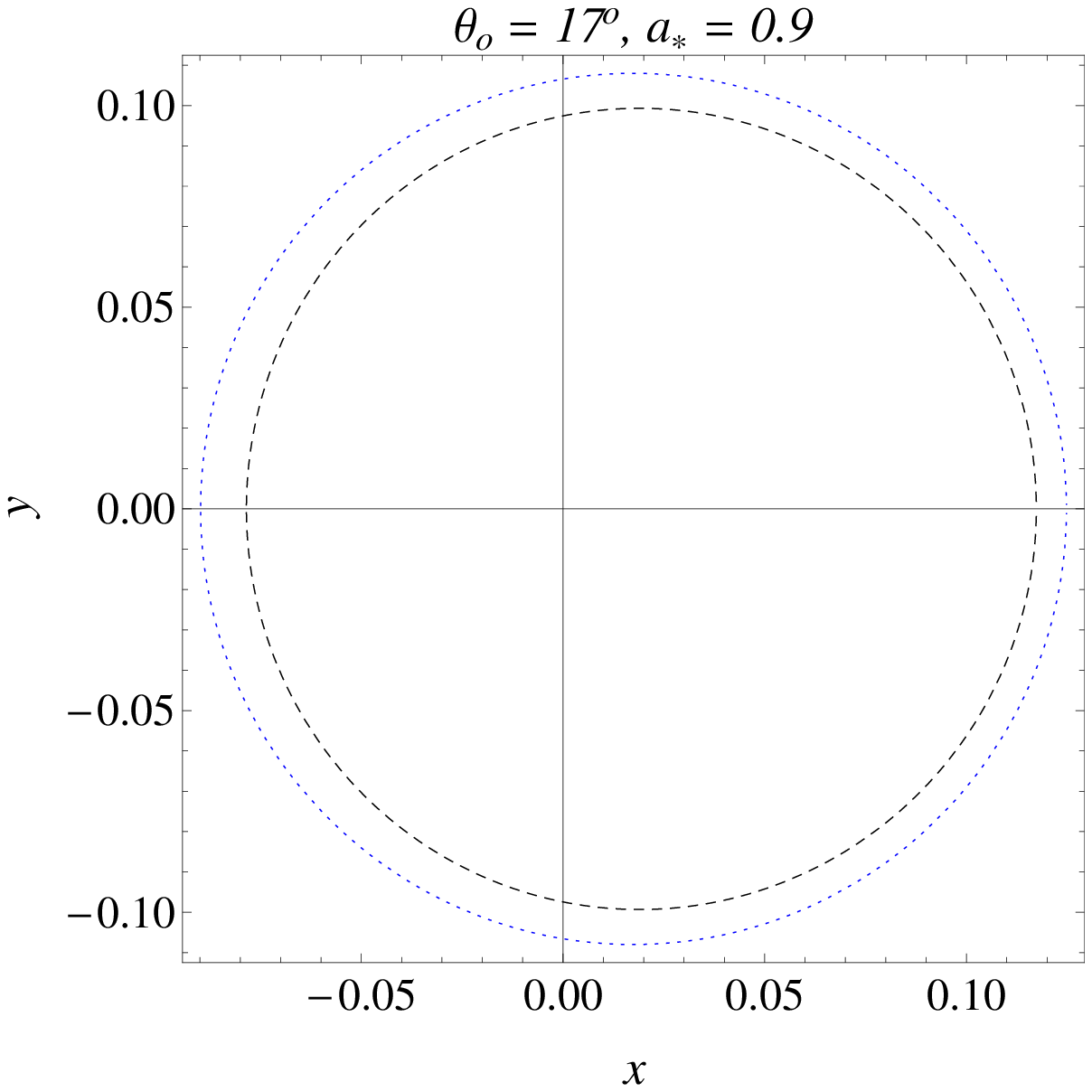}
\includegraphics[scale=0.45]{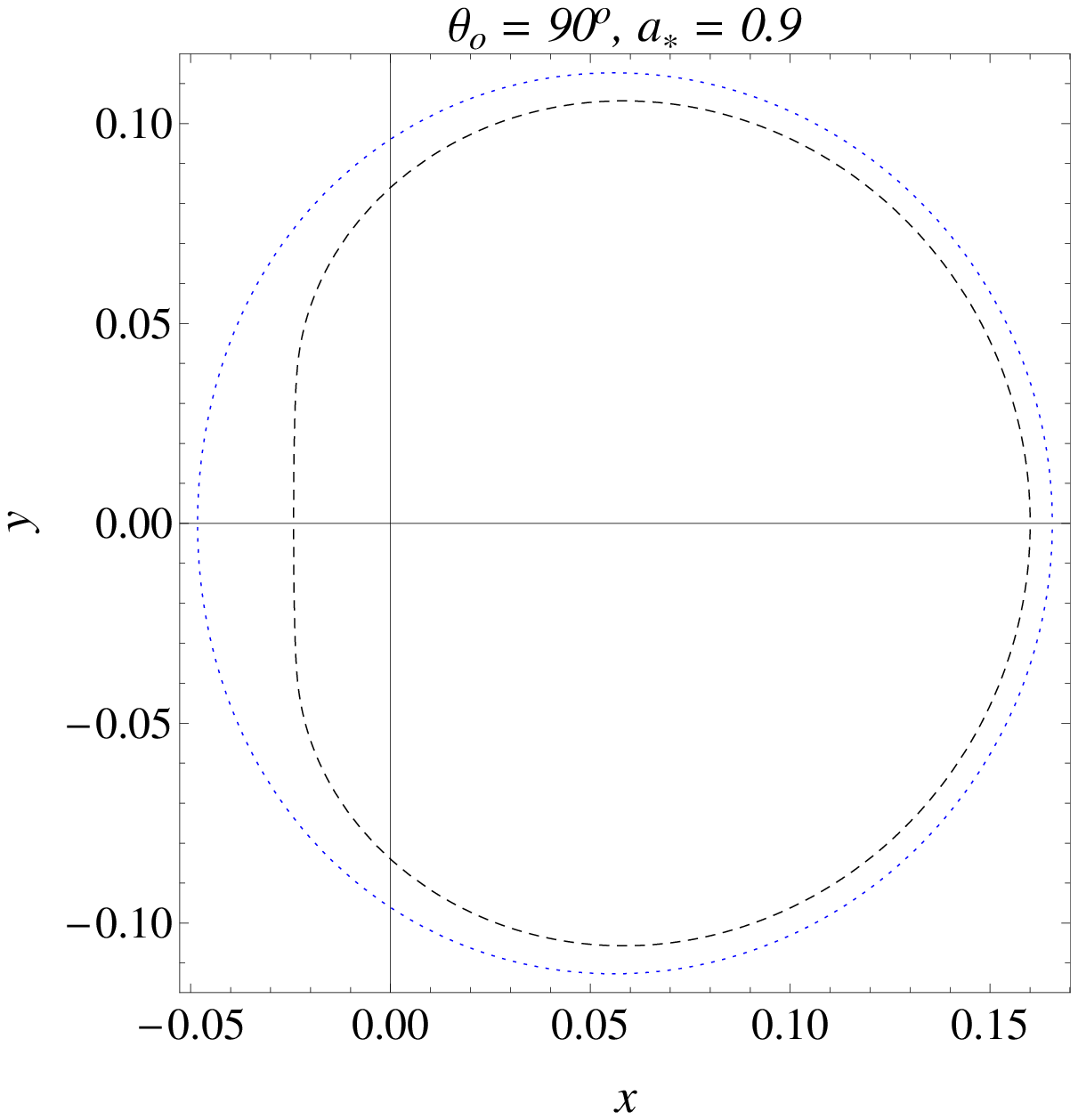}
\par\end{centering}
\caption{Effect of the tidal charge $q$ on the shadow of rotating black holes in a brane world. Shadows of a negative tidal charge are blue ones, and a positive tidal charge produces the black shadows. It is assumed the same absolute value for the tidal charge on the top and on the bottom, namely, $\vert q \vert = 0.19M^2$. As we can see, positive values of $q$ influence strongly the shadow's silhouette. The parameters $a_{*}=a/M$, $\Lambda_{4D}=10^{-3}M$, $r_{obs}=35M$, and $M=1$ are adopted in this figure.}
\label{Shadow1}
\end{figure}

According to Fig. \ref{Shadow1}, the shadow is symmetrical to the $x$-axis. As is well known, for a Kerr black hole, 
the shadow's shape will depend strongly on the rotation parameter $a$ and the observer angle $\theta_o$, 
if we consider an observer at infinity. On the other hand, as we saw, for an asymptotically dS 
  black hole, the observer is in the domain of outer communication, and the silhouette depends on the
  cosmological constant as well. Moreover, as we can see in Fig. \ref{Shadow1}, due to the brane context, the tidal
   charge is another important parameter that influences the shadow.

Shadows for the metric (\ref{Boyer-Lindquist}) were widely built in Ref. \cite{Eiroa2}. Since the main point
in this article is the tidal charge, let us focus on the influence of $q$ on the shadow by emphasizing the angle $\theta_{o}=17^{\circ}$ reported by the EHT, according to which this is the polar angle between the observed jets along the rotation
axis of M87* and our position. In the cited article of Eiroa and Sendra \cite{Eiroa2} such an information 
was not known yet.           

 According to shadows published in the literature, the more rotation, the more moved in the positive $x$-direction the shadow is. That is, that motion points toward the rotation direction. In particular, by assuming that the rotation is
  from left to right, the shadow's silhouette is  deformed on the left for large values of $a$, 
      $\theta_o$ and, interestingly, for positive values of $q$. The  difference between the left  and the right sides is
       due to the photon orbits and the spacetime dragging of rotating black holes. On the left of the shadow,  
       photons travel in the same direction of the black hole rotation, and on the right they travel
        in the opposite direction. From the geodesic equations, maximal points on the left and on the right in the shadow's 
        silhouette are given by $r_{p-}$ and $r_{p+}$, respectively.

  In Fig \ref{Shadow1}, it is shown just cases in which the rotation parameter is smaller than the black
  hole mass, i.e., $a^2<M^2$, in order to produce shadows compatible with M87* and a spacetime structure like
  (\ref{Structure2}). Notice that Fig. \ref{Shadow1} indicates that negative values of $q$ produces larger shadows than
  positive ones considering the same set of parameters ($M,a,\theta_o,\Lambda_{4D}, \vert q \vert,r_{obs}$). Indeed, positive
   values of the tidal charge
  induces a deformed silhouette, effect emphasized for large angles of observation (Fig. \ref{Shadow1} on the bottom). Large values of negative tidal charges do not change the shadow circularity even for large values of $\theta_o$. Indeed,
  the deviation from circularity approaches zero as $-q$ increases. Therefore, a clear deviation from circularity is just
   due to positive tidal charges, justifying an upper bound from the deviation only for positive values of $q$.
    This conclusion on the effect of positive and negative tidal charges is in agreement with previous works \cite{Maartens,Aliev} in which the effect of positive tidal charges is considered more radical. And the strong
    effect of a large positive tidal charge motivated the authors of \cite{Banerjee} to claim that a negative value of the tidal
     charge is more acceptable for the M87* parameters. However, following those authors, it is not possible to rule out
     positive values for the tidal charge to date.

\section{An upper bound on the tidal charge}
In this section, the observable defined as deviation from circularity is built. Following the EHT \cite{EHT,EHT2}, I will use the deviation from circularity in order to compute an upper bound on the positive tidal charge $q$. 

\subsection{Calculating the deviation from circularity}
Following Refs. \cite{Bambi,Maeda,Tsupko,Neves}, the shadow's deviation from circularity is provided 
by using the root-mean-square of the shadow's radius $l(r_p)$, which is defined as 
\begin{equation}
l(r_p)=\sqrt{(x-x_c)^2+(y-y_c)^2}.
\end{equation}
The point $(x_c,y_c)$ is at the \enquote{center} of the shadow and not at the center of the Cartesian
coordinates. Therefore, $y_c=0$ and  
$x_c=x(r_{p'})$, where $r=r_p'$  ($r_{p-}<r_{p'}<r_{p+}$) is the value of orbits in which the $y$-coordinate
 is maximum, i.e.,  in such orbits the Cartesian coordinate assumes $y_{m}=y(r_{p'})$. The value $y_m$ is obtained 
from Eq. (\ref{y_approx}) and its maximum, i.e., $\frac{dy(r_p)}{dr_{p}}=0$. Due to  the axial symmetry,
 $\pm y_m$ are maximum and minimum values of $y$.

 From the shadow's radius $l(r_p)$, the average radius or root-mean-square reads
\begin{equation}
l_{RMS}=\sqrt{\frac{1}{(r_{p+}-r_{p-})}\int_{r_{p-}}^{r_{p+}}l(r_p)^2 dr_p}.
\end{equation}
According to Refs. \cite{Bambi,Neves}, the desired deviation from circularity is given by 
the root-mean-square distance from the average radius $l_{RMS}$, i.e., it is 
written as\footnote{It is worth emphasizing that 
a different parametrization is adopted here (suggested in Ref. \cite{Neves} and different from
Ref. \cite{Bambi}). In such a parametrization, $(x_R-x_L)$ is in the denominator of Eq. (\ref{DeltaC}).} 
\begin{equation}
\Delta C = \sqrt{\frac{1}{(x_R-x_L)}\int_{r_{p-}}^{r_{p+}}\left (l(r_p) - l_{RMS} \right)^2 dr_p},
\label{DeltaC}
\end{equation}
in which two new points of the shadow's silhouette are adopted: $x_L$ and $x_R$ (the most negative
and the most positive values assumed by the shadow's silhouette on the $x$-axis). According to 
Eq. (\ref{x_approx}), these points are given by $x_L=-\sin\alpha(r_{p-})$ 
and $x_R=\sin \alpha(r_{p+})$, where
 $r_{p-}$ is solution of $\sin \beta=1$, and $r_{p+}$ is solution of $\sin \beta=-1$ (see 
Refs. \cite{Neves,Tsupko} for details).

Some comments on the deviation from circularity $\Delta C$: it increases with the black hole rotation and the observation
  angle (see Fig. \ref{Shadow1}), besides a positive tidal charge also increases $\Delta C$. As mentioned before, 
  the EHT reported $\Delta C \lesssim 0.1$ assuming a Kerr metric as the geometry
   of the M87* black hole. In this work, instead, a rotating black hole in a brane world with a cosmological
   constant is adopted. With an  appropriate value for the tidal charge, it is possible to reproduce the M87* shadow.

\subsection{Constraining the tidal charge}

\begin{figure}
\begin{centering}
\includegraphics[scale=0.6]{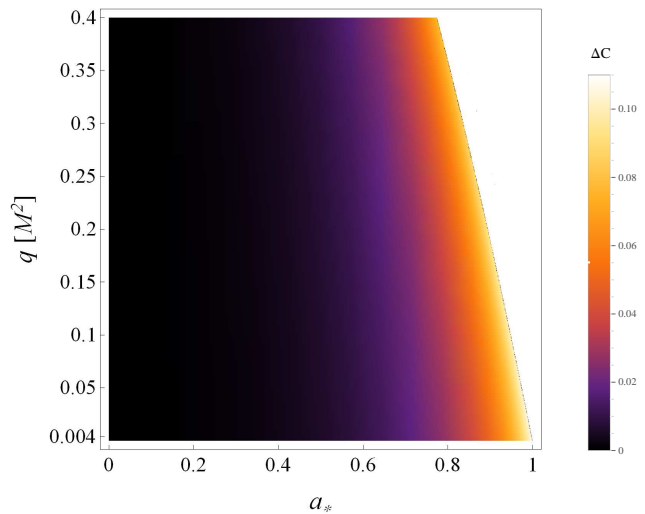}
\par\end{centering}
\caption{Deviation from circularity, $\Delta C$, of the rotating black hole in a brane world and its dependence on
the dimensionless rotation parameter, $a_*=a/M$, and the tidal charge $q$. The white area indicates
forbidden solutions for sets of free parameters ($a_*,q$). The parameters of the graphic are the
M87* parameters  $M=(6.5 \pm 0.7 )\times 10^{9}M_{\odot}$, $\theta_{o}=17^{\circ}$ and
 $r_{obs}= (16.8 \pm 0.8)$Mpc, alongside the cosmological constant $\Lambda_{4D} \simeq 1.1\times 10^{-52}$m.}
\label{Deviation}
\end{figure}

According to Eq. (\ref{DeltaC}), the deviation from circularity depends on the metric parameters. Therefore, 
by using the M87* parameters (namely, $M=(6.5 \pm 0.7 )\times 10^{9}M_{\odot}$ and
 $r_{obs}= (16.8 \pm 0.8)$Mpc) alongside the value of the four-dimensional cosmological constant 
 ($\Lambda_{4D} \simeq 1.1\times 10^{-52}$m), the reported deviation from circularity 
 imposes the following upper bound on the positive tidal charge:
\begin{equation}
\Delta C (a_*,q) \lesssim 0.1  \Rightarrow q  \lesssim 0.004 M^2.
\end{equation}
The above upper bound is obtained for $ a_* \lesssim 0.99$ (with $a_*=a/M$) and $\theta_o=17^{\circ}$ (see
Fig. \ref{Deviation}). 
It is important to note that with $a_* > 0.99$ and the M87* parameters, the spacetime structure with horizons given 
by Eq. (\ref{Structure2}) is not achieved. Thus, for $\theta_o=17^{\circ}$ and the reported deviation, 
the parameter $a_* > 0.99$  is ruled out.
 According to EHT collaboration \cite{EHT}, the rotation parameter of
M87* is still in dispute. However, recent works have constrained the dimensionless rotation parameter of M87*
 by using the observed jets, accretion disk, and light propagating near the black hole: one finds $a_* > 0.4$ in Nemmen \cite{Nemmen} and
 $a_*=0.90 \pm 0.05$ in Tamburini \textit{et al.} \cite{Tamburini}.

 As mentioned before, there exists a maximum value of the tidal charge 
in order to produce the spacetime structure (\ref{Structure2}). Using the M87* parameters such a maximum value is
 $q_{max} \lesssim 0.019 M^2$ for $a_* \lesssim 0.99$. As we can see, the constraint from the shadow 
 improves that value.
 
  \begin{table}
\caption{Normalized values of the positive tidal charge $q$ obtained for some astrophysical objects
 in different approaches. The upper bound obtained here using the M87* parameters is the second best.}
\label{Values of q}
\begin{ruledtabular}
\begin{tabular}{lcc}
     
      & $q/M^2$& \\ \hline
      \text{Sun\footnote{Using light deflection around the Sun \cite{Bohmer}.}} & $\lesssim 0.003$ & \\ 
      \text{M87*}& $\lesssim 0.004$ &  \\ 
      \text{Sagittarius A*I \footnote{Using trajectories of bright stars around the black hole \cite{Zakharov2}.}}&
       $\lesssim 0.4$   & \\
       \text{Sagittarius A* II \footnote{Using gravitational lensing of stars around Sagittarius A* \cite{Bin-Nun}.}}&
        $\lesssim 1.6$   & \\
      \text{Neutron Star Binary\footnote{Using quasiperiodic oscillations in binary systems \cite{Kotrlova}.}}&
       $\lesssim 2.3$  &     \\ 
     \text{Sagittarius A* III \footnote{Using gravitational lensing \cite{Horvath}.}}& $\lesssim 4.5$   & \\
                  
\end{tabular}
\end{ruledtabular}
\end{table}

According to Table \ref{Values of q}, other constraints on the tidal charge 
have been reported in articles.  Contrary to this work, the major part uses a nonrotating geometry in order 
to describe astrophysical black holes and constrain the tidal charge. The best normalized value $q/M^2$ comes 
from light deflection of the Sun  \cite{Bohmer}. A supermassive 
black hole, Sagittarius A*, was also used in \cite{Zakharov2,Bin-Nun,Horvath}, but the normalized upper bound 
reported here is even better.

\section{Final remarks}
The most commented image in science in 2019 was the first image of a black hole announced by the EHT 
collaboration \cite{EHT,EHT2}. According to collaboration, the black hole shadow is well-described
 by the Kerr metric. However, as we saw,  it is argued that options to the Kerr geometry are still possible within the M87*
  parameters. For example,  Bambi \textit{et al.} \cite{Bambi} do not rule out a superspinar or an overspinning 
  black hole  as a candidate for the object that produces the M87* shadow, and Ref. \cite{Neves} 
  suggests a rotating regular black hole as a candidate for M87*. Here, another option is still suggested: 
  a brane black hole with rotation and cosmological constant. 
  
 The brane world context adopted here is the RS-II-type model, in which our four-dimensional universe is immersed
  in a five-dimensional bulk. As is well known in the large literature, the bulk acts nonlocally on the brane, 
  generating a tidal charge even in the empty brane. Using a rotating black hole with a cosmological in the vacuum brane, 
  the tidal charge was constrained  with the aid of the shadow deviation from circularity from the central supermassive 
  black hole  M87* reported by the EHT.  A deviation from circularity $\lesssim 10\%$ indicated 
  a normalized upper bound on the tidal charge $\lesssim 0.004$,  which is the second best result for the
   normalized tidal charge to date (the best one is $\lesssim 0.003$ for a solar system result).

\section*{Acknowledgments}
This study was financed in part by the Coordenação de Aperfeiçoamento de Pessoal de Nível Superior---Brasil (CAPES)---Finance Code 001.

\end{document}